\documentclass[twocolumn]{autart}    

\usepackage{graphicx,amsmath,amssymb,overpic,here,multirow}          
\usepackage{booktabs}
\usepackage{siunitx}

\newtheorem{propos}{\bf Proposition}[section]

\newtheorem{definition}{\bf Definition}[section]

\usepackage[dvipsnames]{xcolor}
\usepackage{pgfplots} 
\pgfplotsset{compat=newest} 
\pgfplotsset{plot coordinates/math parser=false} 
\newlength\figureheight 
\newlength\figurewidth 
\usepackage{tikz}
\usepackage{pgfplots}
\usetikzlibrary{positioning}

\usepackage{algorithm}
\usepackage{algpseudocode}
\begin{document}

\begin{frontmatter}

\title{On Probabilistic Certification of Combined Cancer Therapies Using Strongly Uncertain Models\thanksref{footnoteinfo}} 

\thanks[footnoteinfo]{Corresponding
author M.~Alamir. Tel. +33476826326. Fax +33476826388.}
\author[GIPSA]{Mazen Alamir}\ead{mazen.alamir@grenoble-inp.fr}    
\address[GIPSA]{{\sc CNRS, GIPSA-lab}, Control Systems Department, University of Grenoble. \\ 11 Rue des Math\'{e}matiques,38400 Saint Martin d'H\`{e}res, France.}
\begin{abstract}
This paper proposes a general framework for probabilistic certification of cancer therapies. The certification is defined in terms of two key issues which are the tumor contraction and the lower admissible bound on the circulating lymphocytes which is viewed as indicator of the patient health. The certification is viewed as the ability to guarantee with a predefined high probability the success of the therapy over a finite horizon despite of the unavoidable high uncertainties affecting the dynamic model that is used to compute the optimal scheduling of drugs injection. The certification paradigm can be viewed as a tool for tuning the treatment parameters and protocols as well as for getting a rational use of limited or expensive drugs. The proposed framework is illustrated using the specific problem of combined immunotherapy/chemotherapy of cancer. 
\end{abstract}
\end{frontmatter}

\section{INTRODUCTION}
The use of dynamic models in the optimization of drug scheduling is nowadays a common practice in academic works. This long tradition involves different paradigms such as optimal control \cite{Swan:88,DePillis:01,Ledzewicz:08,Ledzewicz:2007,Ledzewicz2008295,OCA:OCA793}, predictive control \cite{Chareyron:09}, robust control \cite{alamir_robust_cancer:2014} or nonlinear analytic control design \cite{Kassara2011135,Matveev2002311}. \ \\ \ \\ The dynamic models involved in such studies are typically population models that are built by concatenating functional terms (death rate, transition rates, drug effect terms to cite but few examples). Such models qualitatively capture the main phenomena and represent their strength  and their interaction/coupling through dedicated parameters. \ \\ \ \\ While the qualitative representativity of these models is rather easy to assess, the quantitative matching with  reality strongly depends on the model parameters. The latter are unfortunately unknown for a given patient, are highly dispersed between patients
 and vary with time and during the therapy for a given patient. \ \\ \ \\ 
Some recent works \cite{Kiran2010,Jonsson:2013,alamir_robust_cancer:2014} started attempts to address this issue by using robust design in which the therapy is computed so that some statement can be obtained for a set of parameters rather than for the single nominal parameter vector. A robustness-like statement typically takes the following form: \\ \ \\  
\begin{minipage}{0.01\textwidth}
{\color{gray!50} \rule{1mm}{17mm}}
\end{minipage} 
\begin{minipage}{0.45\textwidth}
{\em The scheduled feedback therapy leads to a predefined tumor contraction for {\sc any} realization of the vector of parameters involved in the model within a predefined bounded set} 
\end{minipage} \ \\ \ \\ \ \\
Therefore, robust design is based on the worst-case analysis and can lead to very conservative/pessimistic design. This is because the worst case is considered no matter how small its probability of occurrence is.  \ \\ \ \\ 
In order to avoid focusing on few unlikely although very bad scenarios, the probabilistic approach seeks statement of the form: 
 \\ \ \\  
\begin{minipage}{0.01\textwidth}
{\color{gray!50} \rule{1mm}{20mm}}
\end{minipage} 
\begin{minipage}{0.45\textwidth}
{\em The scheduled feedback therapy leads to a predefined tumor contraction with a probability no less than $(1-\eta)\%$ over all realizations of the parameter vector assuming that the latter obeys a given probability distribution.} 
\end{minipage} \ \\ \ \\ \ \\
This obviously marginalizes very bad realizations if their probability of occurrence is really small. \ \\ \ \\ 
This paper formalizes this paradigm for the specific case of cancer therapy and gives a complete and understandable instance of it in the specific case of combined therapy of cancer that involves immunotherapy and chemotherapy.\ \\ \ \\ 
It is obvious that given the wide range of problems that can be defined in this context, this paper should be viewed as an introduction to a rich paradigm and a starting point to a large set of variations around the necessary specific formulation adopted in the present paper. \ \\ \ \\ 
The paper is organized as follows: First a general formulation of a class of cancer therapy-related problems is given in section \ref{secformulationtherapy}. Section \ref{secrappelscenarios} recalls the framework and useful results of randomized optimization approach also called the scenario-based approach. The application of this framework to the cancer problem defined in Section \ref{secformulationtherapy} is proposed in section \ref{secappligeneral} in the general case.  Finally, section \ref{secenfin} fully illustrates the previous sections in the particular case of combined immuno/chemotherapy of cancer. The paper ends with Section \ref{secConclusion} that summarizes the paper contribution and gives some hints for future investigation. 
\section{Probabilistic Certification of a Therapy} \label{secformulationtherapy} 
\noindent In this section, the concept of a cancer therapy with probabilistic certification is clearly stated.   
\subsection{The Dynamic Model}
Let us consider a general form of a dynamic system representing the evolution of the tumor and the number of circulating lymphocytes among other necessary quantities under a combined  action of several drugs injection rates $u\in \mathbb{R}^{n_u}$:
\begin{eqnarray}
\dot x=F(x,u,p) \label{eqx} 
\end{eqnarray} 
where $x\in \mathbb{R}^{n}$ is the state of the model while $p\in \mathbb{R}^{n_p}$ stands for the vector of parameters involved in the model. It is assumed in the remainder of the present paper that 
\begin{itemize}
\item $x_1$ stands for the tumor size (to be reduced) 
\item $x_2$ stands for the amount of circulating lymphocytes that is commonly used as an indicator of the patient health/resistance and therefore, any strategy has to be defined such that $C(t)\ge C_{min}$ for all $t\ge 0$.
\end{itemize} 
Other state components may be necessary to describe the model (namely $n\ge 2$) but their exact definition is not needed as far as the presentation of the concepts is concerned. \ \\ \ \\ 
It is assumed that the dynamic model (\ref{eqx}) describes the evolution of the system under the combined effect of $n_u$ different drugs such as chemotherapy, immunotherapy, anti-angiogenesis and so on. 
\subsection{The Feedback-Based Therapy Protocol}  
\begin{figure}
\begin{center}
\includegraphics[width=0.5\textwidth]{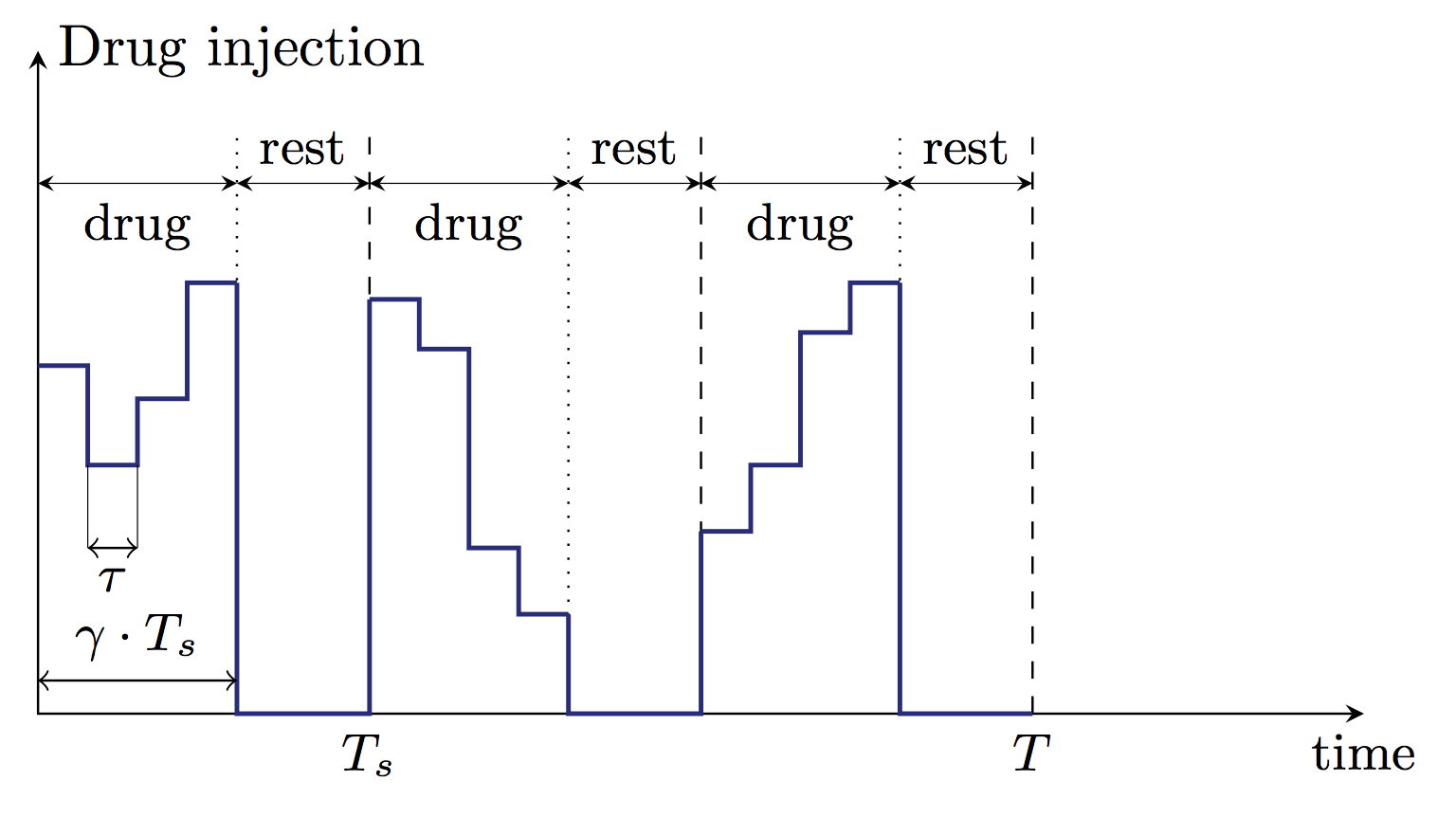}
\end{center} 
\caption{Temporal structure of the therapy. Example of a treatment period consisting of $N_T=3$ sub-periods of a duty cycle $\gamma$.} \label{fig_protocol} 
\end{figure}

\noindent Let us consider a feedback-based therapy of duration $T$ consisting  of $N_T$ sub-periods (of duration $T_s=T/N_T$) each of which involving a treatment phase and a rest phase as shown in Figure \ref{fig_protocol} where the injection curves have to be interpreted as a multivariable signals when several drugs are combined. \ \\ \ \\ 

It is assumed that during a treatment period, a sampled feedback injection law is used with a sampling period $\tau$ (for instance $2$, $4$, $6$ hours or such) during which the injection is maintained constant (see Figure \ref{fig_protocol}):
\begin{eqnarray}
u(k\tau+t)=K(x(k\tau),\theta_c) \qquad t\in [0,\tau] \label{defdeKgen} 
\end{eqnarray} 
where $x(k\tau)$ denote the state of the model at instant $k\tau$ while $\theta_c\in \mathbb{R}^{n_c}$ is a vector of parameters that are used in the definition of the feedback law $K$. \ \\ \ \\ 
In the remainder of the paper, the notation $x(k)$ is used instead of $x(k\tau)$ to simplify the expressions when no ambiguity is possible. It is also assumed that the sampling period is a divisor of $\gamma T_s$ such that there is an integer $N_s$ satisfying:
\begin{eqnarray}
\gamma T_s=N_s\tau \quad;\quad 
\end{eqnarray}  
It is implicitly assumed that the control law (\ref{defdeKgen}) satisfies the following saturation constraints:
\begin{eqnarray}
K_i(x(k),\theta_c)\in [0,u_i^{max}(k)]\qquad i\in \{1,\dots,n_u\} \label{saturation} 
\end{eqnarray} 
where $K_i$ stands for the $i$-th component of $K$ (the $i$-th drug injection value) and where $u_i^{max}(k)$ represents the maximum allowable injection rate of the $i$-th drug during the $k$-th sampling interval. The fact that the maximum injection rate is time-varying is induced by the need to meet the constraint on the total amount of available drugs $D_i$ for the whole therapy. This constraint can be satisfied by using the following definition for $u_i^{max}(k)$: 
\begin{eqnarray}
&&u_i^{max}(k)\le \min\bigl\{\bar u_i,\dfrac{D_i-y_i(k)}{\gamma (T-k\tau)})\\
&&y_i(k+1)=y_i(k)+\tau\times K_i(x(k),\theta_c)\quad;\quad y_i(0)=0 \label{gftYY} 
\end{eqnarray} 
where $y_i(k)$ represents the amount of drug $i$ already injected over $[0,k\tau]$ meaning that $D_i-y_i(k)$ is the available quantity for the remaining therapy duration $T-k\tau$. Note that maximum injection rate is also limited by technical saturation $\bar u_i$ regardless of the amount of initially available drug $D_i$. Note that by definition $\bar u_i$ is not a design parameter since it is imposed by exogenous technical limitations.\ \\ \ \\ 
In section \ref{secenfin}, a fully developed example of such control law is given for the specific example of combined immunotherapy/chemotherapy. For the time being, the general non instantiated form (\ref{defdeKgen}) is kept in order to preserve the general character of the concepts. \ \\ 
\ \\ It comes out that for a given total treatment duration $T$, the therapy is completely defined if the following parameters are defined:
\begin{enumerate}
\item The state feedback parameter vector $\theta_c\in \Theta_c$,
\item The number of sub-periods $N_T$,
\item The duty cycle $\gamma$,
\item The allocated drug quantities $D_i$, $i\in \{1,\dots,n_u\}$
\item The tumor contraction ratio $\gamma_c$ [see (\ref{constr1})] 
\end{enumerate} 
These parameters are gathered in the sequel into a single decision vector $\theta$, namely:
\begin{eqnarray}
\theta= \begin{pmatrix}
\theta_c&N_T&\gamma&\gamma_c&D_1&\dots D_{n_u}
\end{pmatrix}^T \label{gftrtr} 
\end{eqnarray} 
in order to state the probabilistic certification problem discussed in the next section. In the sequel, $\theta$ defined by (\ref{gftrtr}) is referred to as the therapy design parameter while $\theta_c$ is called the feedback design parameter. Therefore, the therapy design parameter set includes the control parameter $\theta_c$ but also the time structure and the maximum injection rates. 
\subsection{The Concept of Probabilistic Certification}
\noindent Note that given the dynamic model (\ref{eqx}), the model's parameter vector $p$ and the therapy parameter vector $\theta$, the evolution of the system can be predicted for any given initial state $x_0$ (at the beginning of the therapy) so that the value of the state $x(k)$ at the sampling instant $k\tau$ can be denoted by:
\begin{eqnarray}
x(k)=:X(k\tau,p,\theta,x_0)
\end{eqnarray} 
Considering a target tumor contraction ratio $\gamma_c$ at the end of the therapy, the success of the therapy can be summarized by the fulfillment of  the following two constraints:
\begin{eqnarray}
&&\dfrac{X_1(T,p,\theta,x_0)}{x_1(0)}\le \gamma_{c} \in [0,1[ \label{constr1}\\
&&\max_{k}\left[C_{min}-X_2(k\tau,p,\theta,x_0)\right]\le 0 \label{constr2} 
\end{eqnarray} 
as this means that the tumor would be contracted by at least $\gamma_c$ at the end of the therapy while the lymphocytes are maintained higher than their lower level $C_{min}$. Obviously, these two constraints can be gathered in a single boolean indicator:
\begin{eqnarray}
g(\theta,p):= \left\{ 
\begin{array}{ll}
 0& \mbox{\rm if (\ref{constr1})-(\ref{constr2}) are satisfied}\\
 1& \mbox{otherwise}
\end{array}
\right. \label{defdeg} 
\end{eqnarray}
where $T$ and $x_0$ are supposed to be given and are therefore omitted from the list of arguments. Since the indicator $g$ is $1$ when the constraints are violated, this indicator is referred to as the failure indicator. 
\begin{definition}[\bf The Failure Indicator]\ \\ 
The function $g(\theta,p)$ defined by (\ref{defdeg}) is called the failure indicator for the therapy defined by $\theta$ and the model defined by the parameter vector $p$. 
\end{definition}
Note however that the constraints (\ref{constr1})-(\ref{constr2}) involve the {\sc unknown} parameter vector $p$. This makes the definition of a successful strategy rather ambiguous. Indeed, two statements are possible as mentioned in the introduction of this paper:
\begin{enumerate}
\item A {\bf Robustly certified} therapy would be the one for which, when using the setting defined by $\theta$, a failure indicator $g(\theta,p)=0$ holds for {\sc any} possible realization of $p$ within the admissible set $\mathbb P$. \\ 
\item A {\bf $(\delta,\eta)$-Probabilistically certified} therapy would be the one for which, when using the setting defined by $\theta$ one can state with a probability no less than $1-\delta\approx 1$ that the expectation of the failure indicator $g(\theta,p)$ over $\mathbb P$ (using the probability measure $\mathcal P$) is at most equal to $\eta\approx 0$. 
\end{enumerate} 
For obvious reasons, the parameter $\delta$ is referred to as the confidence parameter (since it is the probability that the success statement is wrong) while $\eta$ is referred to as the precision parameter since it represents the error committed w.r.t the ideal achievement $g=0$. Note that a robustly certified therapy is a $(0,0)$-probabilistically certified therapy. \ \\ \ \\ 
As mentioned in the introduction, the first concept generally leads to very pessimistic design since it is based on the worst case scenario even if it is very unlikely. Moreover the corresponding computation is extremely difficult. The second concept leads to more tractable computation and it neglects very unlikely bad scenarios leading to more pragmatic design. This is the option followed in the remainder of the paper. \ \\ \ \\ 
In the above discussion, only the satisfaction of the constraints (\ref{constr1})-(\ref{constr2}) is considered. As a matter of fact, there might be several values of $\theta$ that meet the constraints in which case the best $\theta$ should be defined through some cost function $J(\theta)$ to be minimized. In our problem this may be 
\begin{itemize}
\item[$\checkmark$] the quantities of the used drugs (which are a part of $\theta$ according to (\ref{gftrtr}))
\item[$\checkmark$] the duty cycle $\gamma$ reducing  hospitalization periods
\item[$\checkmark$] any convex combination of the above indicators. 
\end{itemize} 
In the next section, the computational aspect that enables to compute $\theta$ leading to a probabilistic certification of the therapy is introduced by recalling the main ideas on the general topics of randomized methods \cite{alamo2009randomized,alamo2015}. 
\section{Recalls on Randomized Methods} \label{secrappelscenarios} 
\noindent Consider the following robust optimization problem in the decision variable $\theta\in \Theta\subset \mathbb{R}^{n_\theta}$ and the uncertainty $p$: 
\begin{eqnarray}
\min_{\theta\in \Theta} J(\theta)\quad \mbox{\rm under $\quad (\forall p) \quad g(\theta,p)=0$} \label{gfvc6} 
\end{eqnarray}  
where $g(\theta,p)$ is defined as in (\ref{defdeg}) by:
\begin{eqnarray}
g(\theta,p):= \left\{ 
\begin{array}{ll}
 0& \mbox{\rm if specification are satisfied}\\
 1& \mbox{otherwise}
\end{array}
\right. \label{defdeggen} 
\end{eqnarray}
and where a probability measure $\mathcal P$ is associated to the uncertainty vector $p$ that is assumed to belong to some admissible set $\mathbb P$.\ \\ \ \\ 
The randomized method replaces the original hard problem (\ref{gfvc6}) by the following problem:
\begin{eqnarray}
\min_{\theta\in \Theta} J(\theta)\quad \mbox{\rm under $\quad \mbox{\rm Pr}_\mathcal{P}\{g(\theta,p)=1\}\le \eta $} \label{hgvc76} 
\end{eqnarray} 
where Pr$_\mathcal{P}\{g(\theta,p)=1\}$ represents the probability of the {\em event}  $g(\theta,p)=1$ (violation of the requirement) when $p$ is randomly generated in accordance with the probability measure $\mathcal P$. \ \\ \ \\ Now since the computation of the probability term is a rather involved and expensive task, the randomized method \cite{alamo2009randomized,alamo2015} simplifies (\ref{hgvc76}) by replacing the probability by the mean value over $N$ drawn independent identically distributed (i.i.d) samples of $p$ in $\mathbb P$, namely the new optimization problem becomes:
\begin{eqnarray}
\min_{\theta\in \Theta} J(\theta)\quad \mbox{\rm under $\quad \displaystyle{\sum_{\ell=1}^N} g(\theta,p^{(\ell)})\le m$} \label{hgvc7poiju7} 
\end{eqnarray} 
which simply replaces the constraint on the probability by a different constraint stating that the mean value of $g(\theta,p^{(\ell)})$ over $N$ random trials to be lower than $m/N$, or to state it differently that at most $m$ between the total number $N$ of trials lead to the violation of the specification. It comes therefore that $N$ must be such that: 
\begin{eqnarray}
\dfrac{m}{N}\le \eta \label{msurN} 
\end{eqnarray} 
which is obviously only a necessary condition. This is because $N$ must also be sufficiently large so that the fulfillment of (\ref{hgvc7poiju7}) implies that the  condition (\ref{hgvc76}) on the probability is satisfied with a probability greater than $1-\delta$ with a pre-specified small value $\delta$. that is the reason why the minimum value of $N$ that makes this implication true involves both the precision specified by $\eta$ and the confidence specified  by $\delta$.\ \\ \ \\ 
In \cite{alamo2009randomized,alamo2015}, several expressions for the value of $N$ are given under different assumptions. In this paper, we are interested in the particular case where the set of design parameter $\theta$ is discrete with cardinality $n_\Theta\in \mathbb N$. This is because some of the parameters being involved such as the available quantities of drugs $D_i$, the number of sub-periods $N_T$ are naturally quantified and cannot be viewed as a free real variables. For all the remaining variables, one can take some representative values on the admissible intervals. By doing so, the optimization problem (\ref{hgvc7poiju7}) is greatly simplified since it can be solved by simple enumeration. Obviously, mixed integer nonlinear programming can also be used following the same lines presented in the paper without significant qualitative difference. \ \\ \ \\ 
According to  \cite{alamo2015}, in this case, the following proposition holds:
\begin{propos} \label{gvfPropS} 
Let $m\in \mathbb N$ be any integer. Let $\delta\in (0,1)$ be a targeted confidence parameter and $\eta\in (0,1)$ be a targeted precision parameter. Take $N$ satisfying:
\begin{eqnarray}
N\ge \dfrac{1}{\eta}\left(m+\ln(\dfrac{n_\Theta}{\delta})+\left(2m\ln(\dfrac{n_\Theta}{\delta})\right)^{1/2}\right) \label{defdeNmini} 
\end{eqnarray}  
then any solution to (\ref{hgvc7poiju7}) in which the $\left\{p^{(\ell)}\right\}_{\ell=1}^N$ are randomly i.i.d drawn using the probability measure $\mathcal P$ satisfies the constraint in (\ref{hgvc76}) with a probability $\ge 1-\delta$.  
\end{propos}
A remarkable property of the expression (\ref{defdeNmini}) enabling the computation of $N$ is that it is totally independent of the the dimension of $p$ (the number of parameters involved in the dynamic model in our application). This is of tremendous importance in the context of certified therapy since there are typically a quite high number of parameters (these are typically the gains associated to each functional term in the model). It is a rather good news that this does not influence the number of trials that is needed to define the constraint in the optimization problem (\ref{hgvc7poiju7}). This is a rather counter intuitive feature for a simple first thought.  \ \\ \ \\ 
Another interesting feature of Proposition \ref{gvfPropS} is that the confidence parameter $\delta$ appears through a logarithmique term which means that one can seek highly confident assertions without dramatic increase in the number of trials. \ \\ \ \\ 
In the next section, the use of the randomized method summarized in Proposition \ref{gvfPropS} in the certification of cancer therapy is presented in the general setting before a specific and complete study of a particular case is proposed in section \ref{secenfin}.
\section{Application to Certification of Therapies} \label{secappligeneral} 
\noindent The application of the framework of the preceding section to the certification of cancer therapy can be achieved using the following setps:
\begin{enumerate}
\item {\bf Definition of the feedback law:} \\ First of all, a state feedback law of the form (\ref{defdeKgen}) has to be designed. This design is problem-dependent although some works seek general structures for the solution such as in \cite{Jonsson:2014}. Another option is to adopt systematic use of generic approaches such as Model Predictive Control (MPC) \cite{Chareyron:09,Mayne:2000} which can be systematically applied as soon as some dynamic model (including potentially nonlinear complex models), a cost function and a constraint function are clearly defined which is the case in our context. Note that free open-source available softwares are now available for practitioners that enable easy implementation of MPC controllers \cite{Houska2011a}. nevertheless, the definition of the control law delivered by such tools still need some parameters to be fixed by the user such as the weighting matrices, the constraints (total drug available), the sampling period and so on. These parameters together with those needed to define the time structure of the therapy defined in Figure \ref{fig_protocol} represent what is referred to in the previous section by the therapy parameter vector $\theta$. \\
\item {\bf Definition of the probability measure $\mathcal P$}\\ 
The feedback law invoked in the preceding item accepts the parameter vector $p$ used in the definition of the model (\ref{eqx}) as a given parameter. As mentioned in the previous section, the use of the randomized method need a probability measure $\mathcal P$ to be defined for use in the generation of the $N$ i.i.d  set of trials $p^{(\ell)}$, $\ell=1,\dots,N$ invoked in the definition (\ref{hgvc7poiju7}) of the relaxed optimization problem. Three main options are here available:\\
\begin{enumerate}
\item In the first, a nominal values $p^{nom}\in \mathbb{R}^{n_p}$ can be used and the probability measure can be defined by a Gaussian distribution around this nominal value with a predefined covariance matrix.  \\
\item Another possibility is to consider that each component $p_i$ belongs to some interval $[\underline p_i,\bar p_i]$ and the probability measure represents a simple uniform distribution (all the values inside the interval are treated with equal probability). \\
\item The last option combines the two preceding one by adopting Gaussian distributions that are saturated by some extreme values $\underline p_i$ and $\bar p_i$ in order to avoid unrealistic trials to take place (such as negative values for a intrinsecly positive parameter). \\
\end{enumerate}   
\item {\bf Definition of the Confidence and precision parameters}. \\ These are the parameters $\delta$ and $\eta$ involved in the  randomized approach. Recall that $1-\delta$ defines the confidence with which the certification result can be assessed while $1-\eta$ represents the probability of failure in the fulfillment of the constraints. Typical values for these parameters are $\delta=10^{-3}$ and $\eta=10^{-2}$.\\
\item {\bf Definition of the design parameter set $\Theta$}.\\ This is done by choosing for each component $\theta_i$ of the therapy design  parameter a set of representative values 
\begin{eqnarray}
\Theta_i:=\{\theta_i^{(1)},\dots,\theta_i^{(n_i)}\}
\end{eqnarray} 
covering the presumed interval of interesting values. This obviously leads to a discrete set of cardinality:
\begin{eqnarray}
n_\Theta=\prod_{i=1}^{n_\theta}n_i\qquad;\qquad \Theta=\prod_{i=1}^{n_\theta} \Theta_i
\end{eqnarray} 
recall that the impact of the value of $n_\Theta$ on the complexity of the solution (through the number of trials $N$) appear through a logarithm which means that high values of $n_\Theta$ can be used to reasonably explore the design space. \ \\ \ \\ 
We assume that a numbering rule is used inside $\Theta$ so that the $n_\Theta$ elements of the discrete set $\Theta$ can be denoted as follows:
\begin{eqnarray}
\Theta:=\Bigl\{\theta^{(\sigma)}\Bigr\}_{\sigma=1}^{n_\Theta}\quad;\quad \theta^{(\sigma)}\in \mathbb{R}^{n_\theta}
\end{eqnarray} 
\item {\bf Computing the sample size $N$}. This can be done by choosing an arbitrary value of $m$ (say $m=1$) and using the above mentioned $\delta$, $\eta$ and $n_\Theta$ in (\ref{defdeNmini}) to compute $N$.  Table \ref{lesNtab} shows the evolution of the sample size $N$ (number of trials needed to achieve the certification) as a function of the precision $\eta$ and the cardinality $n_\Theta$ of the design parameter set $\Theta$. The confidence parameter is systematically taken equal to $\delta=10^{-3}$.
\begin{table}
\begin{center}
\begin{tabular}{lcccc} \toprule
    {$n_\Theta$} & {$\eta=0.1$} & {$\eta=0.05$} & {$\eta=0.01$} & {$\eta=0.001$} \\ \midrule
    1  & 132 &       264 &      1317 &       13164 \\
    5  & 154 &       308  &     1536    &    15354\\
    10  & 163 &        326 &         1628    &    16280\\
    100  & 193 &       386 &       1930  &      19299\\
    1000  & 223 &       445 &       2225 &       22249\\
    10000  &  252  &     503   &    2515    &    25148\\
 \bottomrule
\end{tabular}
\end{center} 
\ \\
\caption{Evolution of the sample size $N$ (number of trials needed to achieve the certification) as a function of the precision $\eta$ and the cardinality $n_\Theta$ of the design parameter set $\Theta$ (confidence parameter $\delta=10^{-3}$ is used).} \label{lesNtab} 
\end{table}
\ \\
\item {\bf Draw the model parameter samples}. \\ 
Having $N$ at hand, a set of $N$ sample $p^{(\ell)}$, $\ell=1,\dots,N$ is drawn using the probability measure defined in step  (2) above. \\ 
\item {\bf Perform Closed-loop simulations}. \\
In this step, for each of the $n_\Theta$ candidate values $\theta^{(\sigma)}\in \Theta$, $\sigma=1,\dots,n_\Theta$ defined in step (4) and each of the model parameter vector $p^{(\ell)}$ generated in step (6), the resulting model (with $p^{(\ell)}$ used for $p$ in (\ref{eqx})) is simulated using the feedback therapy defined by $\theta^{(\sigma)}$). This obviously results in $N\cdot n_\Theta$ closed-loop simulation of the model over the therapy duration. Table \ref{lesNtab} can be used to evaluate this number by multiplying each element of the inside matrix (given $N$) by the corresponding line value of $n_\Theta$. For instance, when using $\eta=0.01$ and $n_\Theta=10000$, one needs $2515\times 10000\approx 25\times 10^6$ closed-loop simulations of the therapy.  Note however than with nowadays computers, a single simulation of commonly used population models takes no more than a hundred of microseconds which brings the computation time (even for the very demanding precision level corresponding to $\eta=10^{-2}$ to less than one hour.\ \\ \ \\ 
\begin{rem}
Note that this estimation of the computational task is pessimistic since the candidate values $\theta^{(\sigma})$ can be visited in a clever way so that necessarily unsuccessful values are never tried. For instance, if for some quantity of drugs $D_i$ and a given set of other parameter is unsuccessful, there is no need to visit all those combinaison of parameter that correspond to lower values of $D_i$. \ \\ \ \\ 
\end{rem}
Note that for each simulation corresponding to $(\theta^{(\sigma)},p^{(\ell)})$, the resulting failure indicator: 
\begin{eqnarray}
g^{(\sigma,\ell)}:=g(\theta^{(\sigma)},p^{(\ell)}) \label{defdegsigmaell} 
\end{eqnarray} 
can be computed where $g(\cdot,\cdot)$ is defined by (\ref{defdeg}). Similarly for any candidate cost function (quantity of drugs, duty cycle, etc) the corresponding cost matrix $J^{(\sigma)}$ can be computed. \\
\item {\bf Computing the admissible set of design parameters}\\
Having computed $g^{(\sigma,\ell)}$, the constraints in (\ref{hgvc7poiju7}) can now be evaluated for each candidate parameter $\theta^{(\sigma)}$ by summing the columns of the $\sigma$-th line of the matrix $g^{(\sigma,\ell)}$, namely:
\begin{eqnarray}
\displaystyle{\sum_{\ell=1}^N} g(\theta,p^{(\ell)})=\sum_{\ell=1}^Ng^{(\sigma,\ell)}
\end{eqnarray} 
if the result is lower than $m$ then the candidate value $\theta^{(\sigma)}$ is considered to be admissible. Therefore the admissible set of design parameters is defined by:
\begin{eqnarray}
\mathcal A:=\Bigl\{\sigma\in \{1,\dots,n_\Theta\}\quad \vert \quad \sum_{\ell=1}^Ng^{(\sigma,\ell)}\le m\Bigr\}
\end{eqnarray} 
\item {\bf Compute the optimal certified therapy}   \\
The optimal therapy is defined by $\theta^{(\sigma^*)}$ where $\sigma^*$ is the index of the admissible therapy that minimizes the cost function, namely:
\begin{eqnarray}
\sigma^*= \mbox{\rm arg}\min_{\sigma\in \mathcal A}\left[J^{(\sigma)}\right]
\end{eqnarray}          
\end{enumerate} 
In the next section, the road map detailed above is applied to the specific example of combined immunotherapy/chemotherapy of  cancer.
\section{Illustrative example: Combined immuno/chemo therapy of cancer} \label{secenfin} 
\noindent  The main objective of this example is to enhance a complete understanding of the proposed framework so that future works can be initiated using various models, combination of drugs, cost functions and so on. 
\subsection{The dynamic model}
Consider the dynamic model used in \cite{Kassara2011135} in which a external source of effector-immune cells can be administered in addition to the chemotherapy drugs. The model involves $7$ states and $2$ control inputs that are defined as follows:
\begin{tabbing}
\hskip 1cm \= \hskip 6cm \kill 
$x_1$ \> tumor cell population \\
$x_2$ \> circulating lymphocytes population\\
$x_3$ \> chemotherapy drug concentration \\
$x_4$ \> effector immune cell population \\
$x_5$ \> quantity of already delivered chemo drug\\
$x_6$ \> quantity of already delivered immuno drug\\
$x_7$ \> remaining time for therapy\\
$u_1$ \> rate of introduction of immune cells \\
$u_2$ \> rate of introduction of chemotherapy 
\end{tabbing}
The dynamic model takes the standard form (\ref{eqx}):
\begin{eqnarray}
\dot x_1&=&ax_1(1-bx_1)-c_1x_4x_1-k_3x_3x_1 \label{model1} \\
\dot x_2&=&-\delta x_2-k_2x_3x_2+s_2 \label{model2} \\
\dot x_3&=&-\gamma_0 x_3+u_2 \label{model3} \\
\dot x_4&=&g\dfrac{x_1}{h+x_1}x_4-rx_4-p_0x_4x_1-
k_1x_4 x_3+s_1u_1 \label{model4} \\
\dot x_5&=&u_1 \qquad;\qquad x_5(0)=0\label{model5}\\ 
\dot x_6&=&u_2 \qquad;\qquad x_6(0)=0\label{model6}\\
\dot x_7&=&-1\qquad;\qquad x_5(0)=T\label{model7} 
\end{eqnarray}  
where the description of the role of each groups of term is given in Table \ref{tab1}. Note that the dynamic model (\ref{model1})-(\ref{model4}) involves $n_p=14$ parameters. The nominal values of these parameters as used in \cite{Kassara2011135} are summarized in Table \ref{tab2}.  
\begin{table}
\begin{center}
\begin{tabular}{lll} \toprule
    {Eq.} & Term & Description \\ \midrule
    (\ref{model1}) & $ax_1(1-bx_1)$ & {\footnotesize Logistic tumor growth}\\    
    (\ref{model1})  & $-c_1x_4x_1$ & {\footnotesize Death of tumor due to effector cells}\\
    (\ref{model1})  & $-k_3x_3x_1$ & {\footnotesize Death of tumor due to chemotherapy}\\
    (\ref{model2})  & $-\delta x_2$ & {\footnotesize Death of circulating lymphocytes}\\
(\ref{model2})   & $-k_2x_3x_2$ & {\footnotesize Death of lymphocytes due to chemo}\\
(\ref{model2})   & $s_2$ & {\footnotesize Constant source of lymphocytes}\\
(\ref{model3})   & $-\gamma_0 x_3$ & {\footnotesize Exponential decay of chemotherapy}\\    
(\ref{model4})   & $g\dfrac{x_1}{h+x_1}x_4$ & {\footnotesize Stimulation of tumor on effector cells}\\
(\ref{model4})   & $-rx_4$ & {\footnotesize Death of effector cells}\\
(\ref{model4})   & $-p_0x_4x_1$ & {\footnotesize Inactivation of effector cells by tumor}\\    
(\ref{model4})   & $-k_1x_4x_3$ & {\footnotesize Death of effector cells due to chemo}\\    
\bottomrule
\end{tabular}
\end{center} 
\ \\
\caption{Signification of the terms involved in the dynamic model (\ref{model1})-(\ref{model4}) [source \cite{Kassara2011135}]} \label{tab1} 
\end{table} 
\begin{table*}
\begin{center}
\begin{tabular}{llllll} \toprule
    param & value & param & value  & param & value\\   \midrule  
    $a$ & $4.31\times 10^{-3}\ day^{-1}$ & $b$ & $1.02\times 10^{-14}\ cell^{-1}$ & $c_1$ & $3.41\times 10^{-10}\ (cell\cdot day)^{-1}$\\
$f$ & $4.12\times 10^{-2}\ day^{-1}$ & $g$ & $1.5\times 10^{-2}\ day^{-1}$ & $h$& $2.02\times 10^1\ cell^2$ \\
$k_2,k_3$ & $6\times 10^{-1}\ day^{-1}$ & $k_1$ & $8\times 10^{-1}\ day^{-1}$ & $p_0$ & $2\times 10^{-11}\ (cell\cdot day)^{-1}$\\
$s_1$ & $1.2\times 10^4\ cell\cdot day^{-1}$ & $s_2$ & $7.5\times 10^8\ cell\cdot day^{-1}$ & $\delta$ & $1.2\times 10^{-2}\ day^{-1}$ \\
$\gamma$ & $0\times 10^{-1}\ day^{-1}$ & & & & \\
\bottomrule
\end{tabular}
\end{center} 
\ \\
\caption{Nominal values of the $14$ parameters involved in the dynamic model (\ref{model1})-(\ref{model7}).} \label{tab2} 
\end{table*} 
\subsection{Definition of the feedback law}
\begin{figure}
\begin{center}
\includegraphics[width=0.5\textwidth]{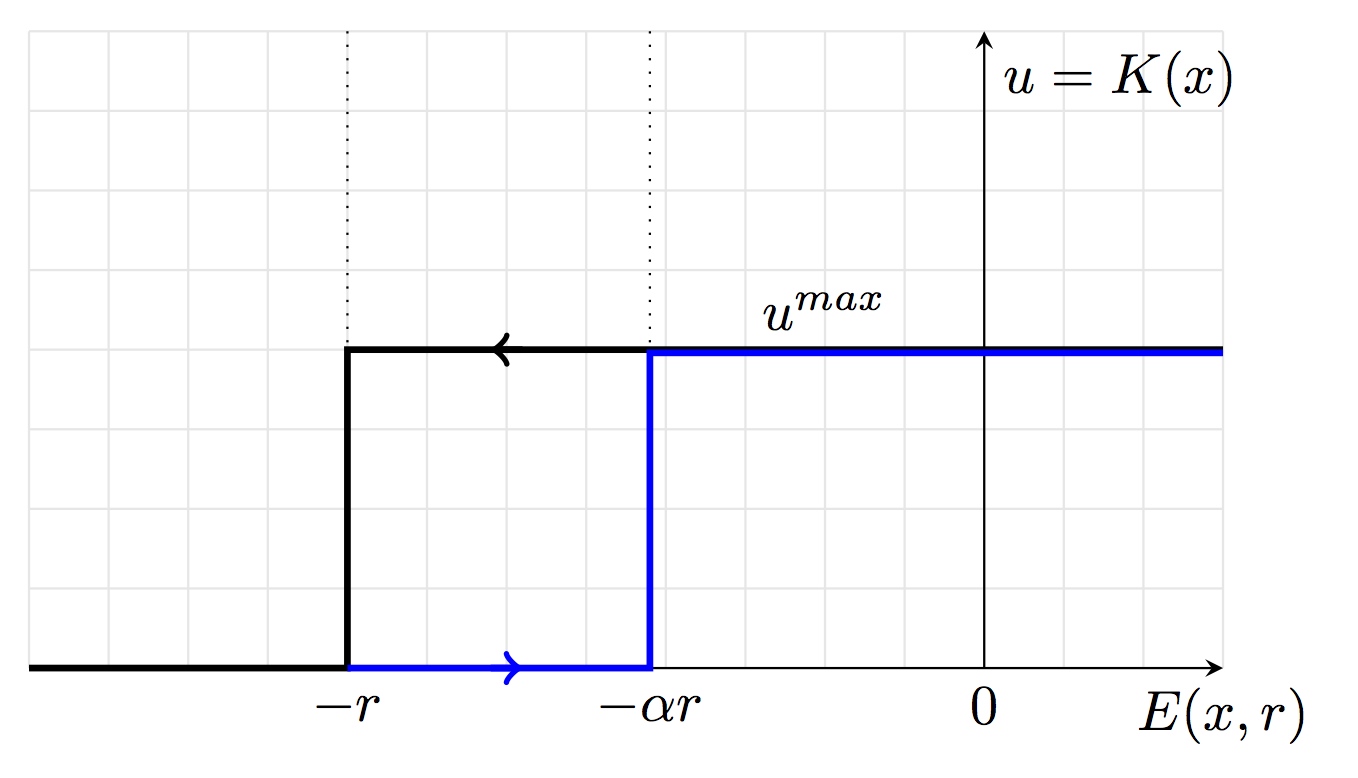} 
\end{center} 
\caption{The definition of the feedback law. Note that $u^{max}\in \mathbb{R}^{2}$ so that the curves have to be interpreted component-wise where $u^{max}$ is given by (\ref{defdeu1max})-(\ref{defdeu2max}).} \label{fighysteresis} 
\end{figure}

\begin{figure}[H]
\begin{center}
\includegraphics[width=0.45\textwidth]{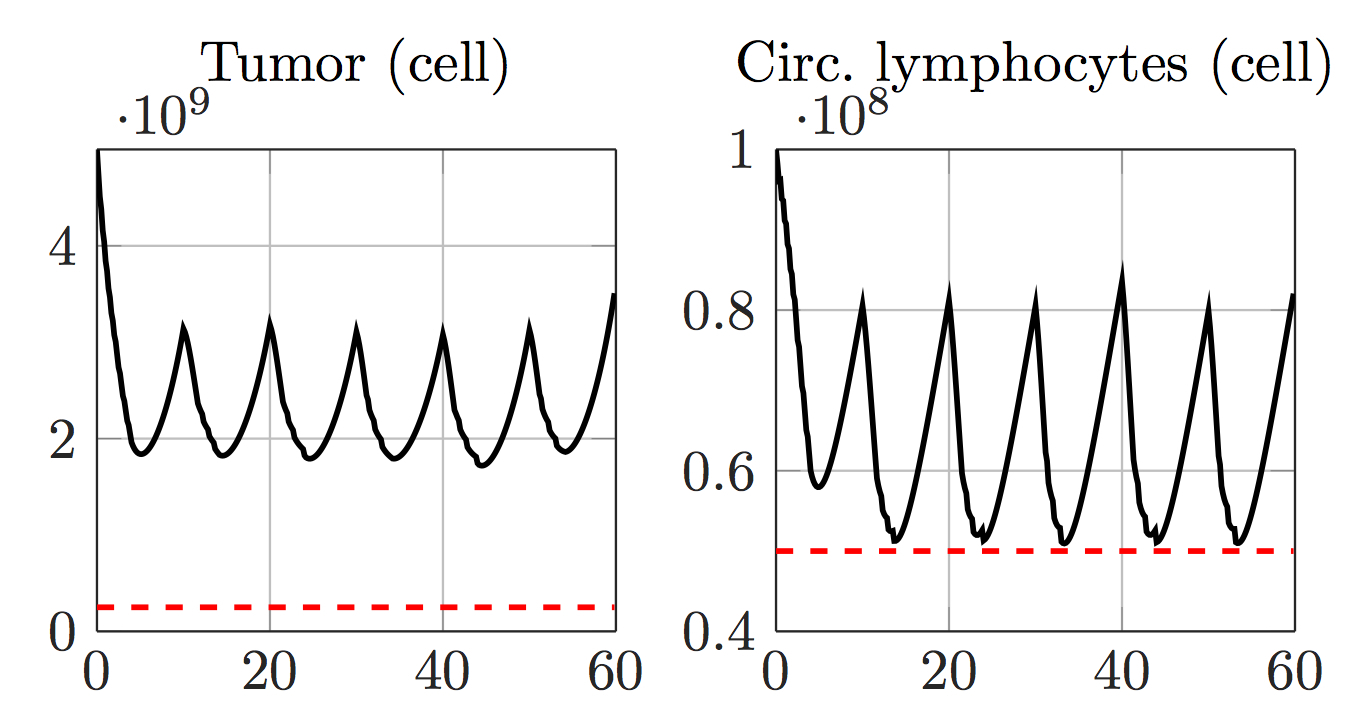} 
\end{center}
{\footnotesize {\bf Simulation 1}. Results with the design parameters $r=0.5$, $D_i=0.75\gamma T \bar u_i$, $\gamma=0.4$, $\beta=1.05$, $N_T=6$ and $\alpha=0.5$}
\begin{center}
\includegraphics[width=0.45\textwidth]{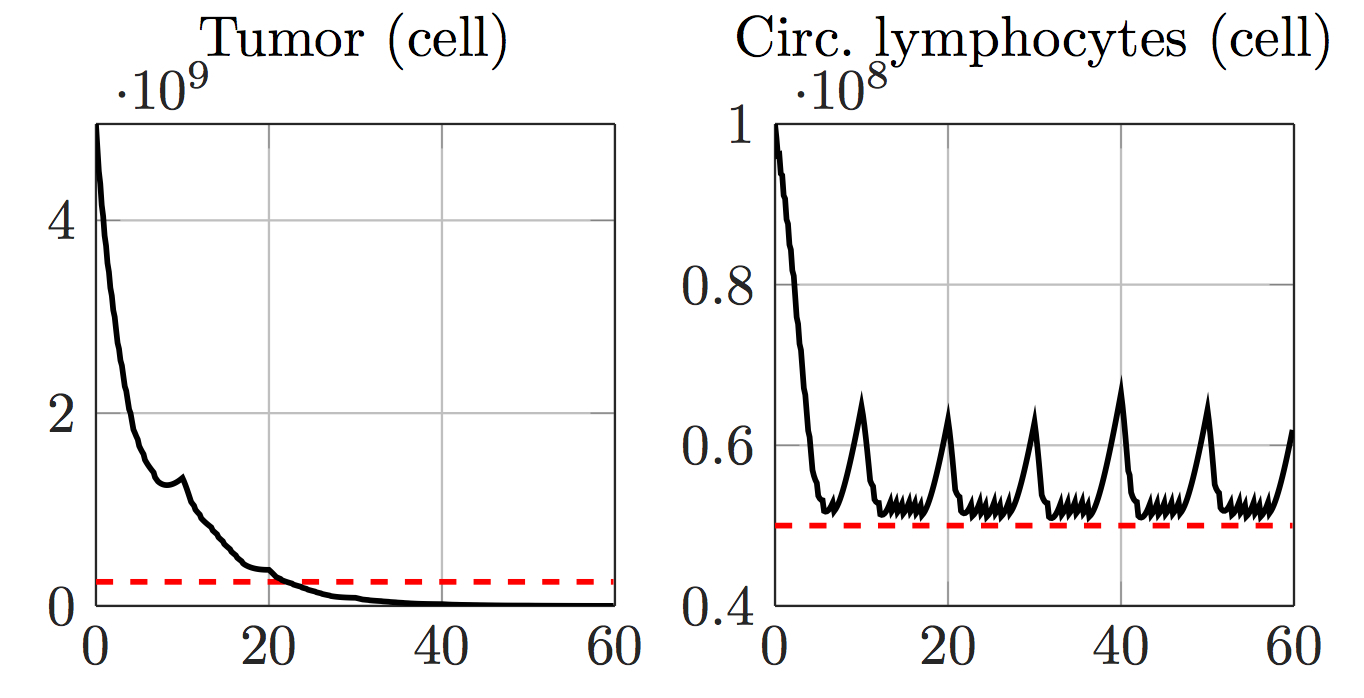} 
\end{center} 
{\footnotesize {\bf Simulation 2}. Results with the same parameters as in Simulation 1 but with the duty cycle $\gamma=0.7$ instead of $0.4$.}
\begin{center}
\includegraphics[width=0.45\textwidth]{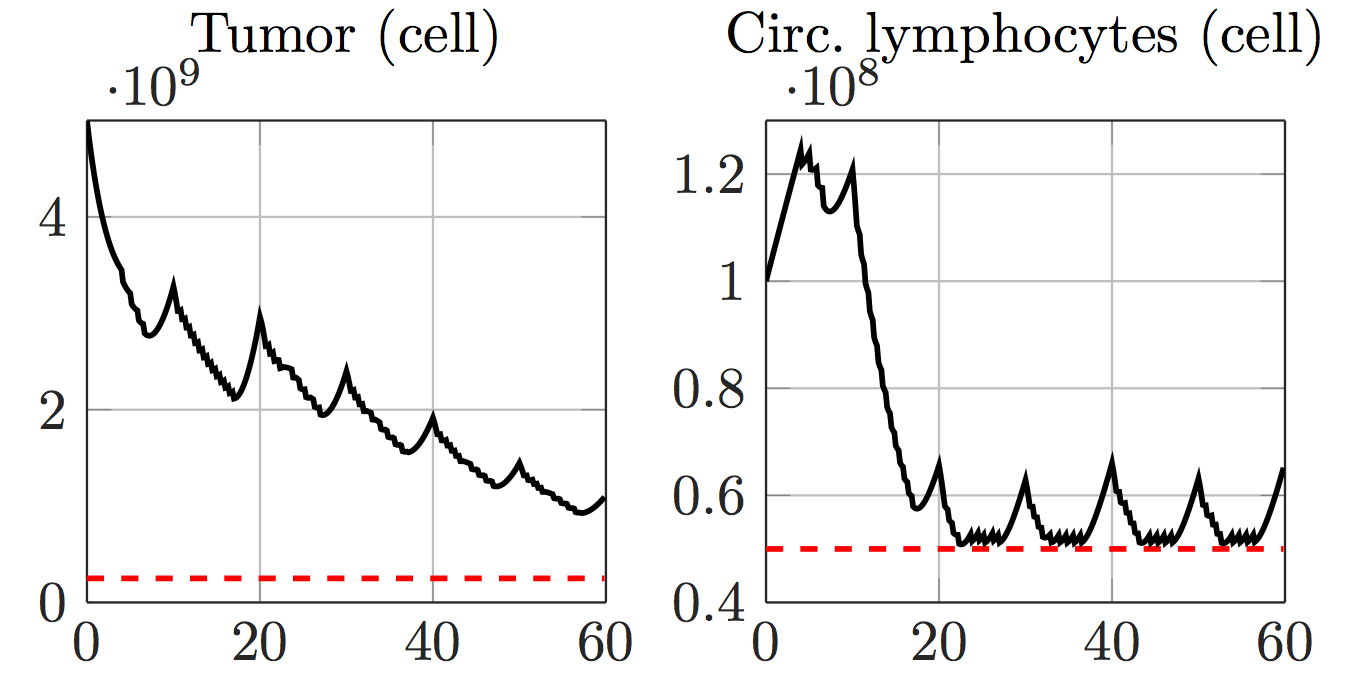} 
\end{center} 
{\footnotesize {\bf Simulation 3}. Results with the same parameters as in Simulation 2 but with the hysteresis parameter $\alpha=0.1$ instead of $0.5$.}
\begin{center}
\includegraphics[width=0.45\textwidth]{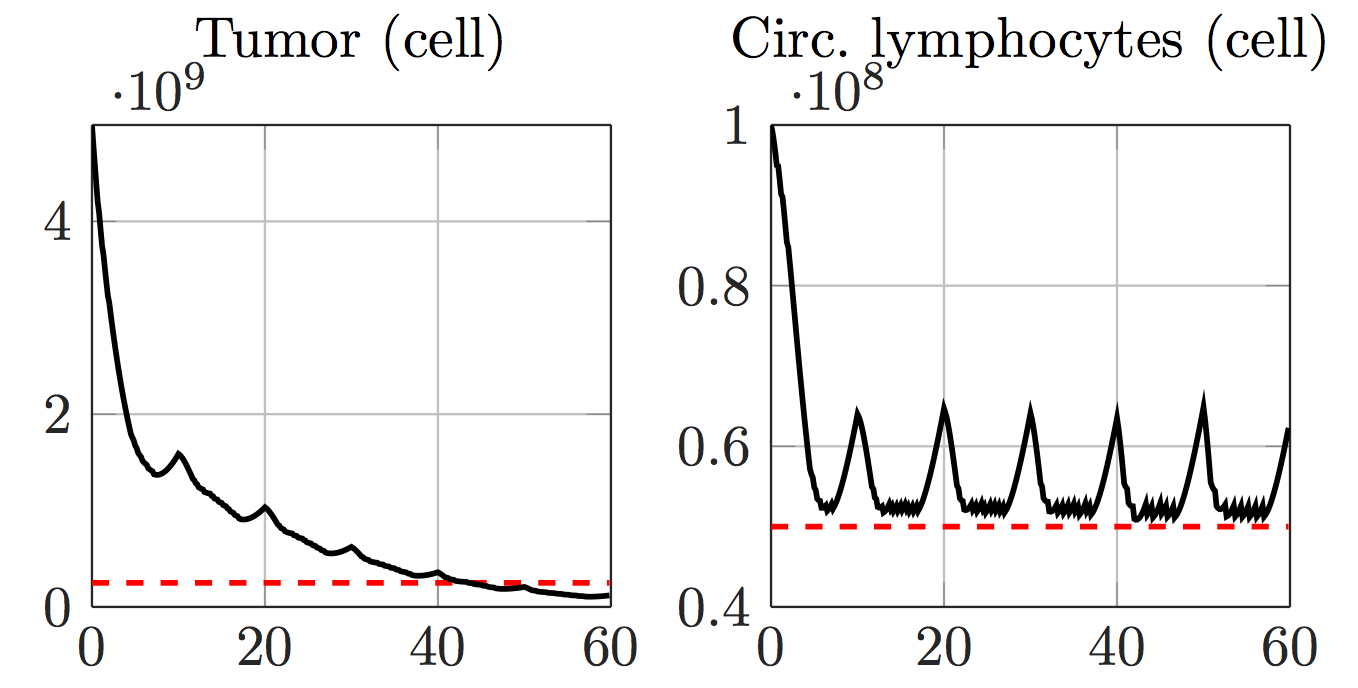} 
\end{center} 
{\footnotesize {\bf Simulation 4}. Results with the same parameters as in Simulation 2 but with the available drug quantities $D_i=0.5\gamma T \bar u_i$ instead of $D_i=0.75\gamma T \bar u_i$}
\caption{Four different simulated therapies with four different sets of therapy design parameter vector $\theta$.} \label{quatrescenaris} 
\end{figure}

\noindent The starting point in the design of the feedback law lies in the fact that according to (\ref{model3}) if one can guarantee that $x_3$ always satisfies the inequality 
\begin{eqnarray}
x_3&\ge& x_3^{max}(x,\beta)\nonumber \\ &:=& 
\max\left\{0,\dfrac{\mu_2(\beta C_{min}-x_2)+\delta x_2-s_2}{-k_2x_2}\right\}\label{defdex3max} 
\end{eqnarray} 
for some $\beta>1$ then according to (\ref{model2}), the evolution of the  lymphocytes population would satisfy the inequality:
\begin{eqnarray}
\dot x_2\ge \mu_2(\beta C_{min}-x_2)
\end{eqnarray} 
which simply would imply that as soon as $x_2$ becomes lower than $\beta C_{min}>C_{min}$ then it can only increase. This obviously prevent the health constraint $x_2\ge C_{min}$ from being violated. \ \\ \ \\ 
The next step is to observe that meeting the inequality (\ref{defdex3max}) on $x_3$ can be guaranteed if one uses equation (\ref{model3}) to induce a corresponding limitation in the chemotherapy drug delivery. This can be done by if the following constraint is satisfied on the chemotherapy drug injection rate $u_2$:
\begin{eqnarray}
u_2\le \gamma_0 x_3^{max}(x,\beta) 
\end{eqnarray} 
This constraint has to be combined with the other constraints (\ref{gftYY}) imposed on $u_1$ in order to meet the technical constraint $u_1\le \bar u_1$ and the one induced by the limited amount of chemotherapy drug that is available for the  therapy. This leads to the following definition of $u_2^{max}$:
\begin{eqnarray}
u_2^{max}(x,\beta)=\min\Bigl\{\bar u_2,\gamma_0 x_3^{max}(x,\beta),\dfrac{D_2-x_6}{\gamma_0 x_7}\Bigr\} \label{defdeu1max} 
\end{eqnarray} 
where the last term comes from (\ref{gftYY}) in which, the already injected chemo drug $x_5$ and the remaining time for the therapy $x_7$ are used. As for the immunotherapy drug, the following simple definition is used since no other limitations are to be considered:
\begin{eqnarray}
u_1^{max}=\min\left\{\bar u_1,\dfrac{D_1-x_5}{\gamma_0x_7}\right\} \label{defdeu2max} 
\end{eqnarray}  
From now on, when we write $u=u^{max}(x,\beta)$, this is to be interpreted component-wise using (\ref{defdeu1max})-(\ref{defdeu2max}). Note that the above definitions involve the first parameter $\beta>1$ that is a part of the control design parameter $\theta_c$. \ \\ \ \\ 
The bounds defined above gives the maximum values that are possible to be administered. The effectively applied values are defined according to the targeted tumor decrease. More precisely, assume that an exponential decrease is targeted. Such decrease would be characterized by the following condition:
\begin{eqnarray}
\dfrac{\dot x_1}{x_1}\le -r \label{defder} 
\end{eqnarray} 
Now for a continuous decrease, when $r=3/T$ one can achieve a settling time (at $0.05$ of the initial value) at the end of the therapy. In the proposed therapy protocol however, because of the potential rest period, much higher value of $r$ would be necessary to achieve the same contraction and this value strongly depends on the drug delivery periods $T_s$ and the duty cycle $\gamma$. That is the reason why $r$ is supposed to be the second component in the control design vector $\theta_c$. \ \\ \ \\ 
Denoting by $F_1(x)$ the r.h.s of (\ref{model1}), the {\em ideal condition} (\ref{defder}) becomes:
\begin{eqnarray}
E(x,r):=\dfrac{F_1(x)}{x_1}\le -r \label{defdeExr} 
\end{eqnarray}   
The idea is then to define the feedback using the hysteresis that is defined in terms of the function $E(x,r)$ as shown in Figure \ref{fighysteresis}. More precisely, the feedback is defined in terms of $E(x(k),r)$ and its past value (over the past sampling period, namely $E(x(k-1),r)$ by:
\begin{eqnarray}
&&\mbox{\rm {\bf If} $x_1>$ threshold} \label{defdethreshold} \\ 
&&u:=\left\{ 
\begin{array}{ll}
u^{max} & \mbox{\rm if $E\ge -\alpha r$}\\
0 & \mbox{if $E\le -r$}\\
u^{max} & \mbox{if $E\in (-r,-\alpha r)$ and $\Delta E< 0$}\\
0 & \mbox{if $E\in (-r,-\alpha r)$ and $\Delta E\ge 0$}\\
\end{array}
\right. \\
&& \mbox{\rm {\bf else} $u=0$}  
\end{eqnarray} 
where $\Delta E(k):= E(x(k),r)-E(x(k-1),r)$ and where $\alpha\in (0,1)$ is a design parameter (this is the third parameter defined so far as a component of the control design parameter vector $\theta_c$. \ \\ \ \\ 
The rational behind this definition can be understood based on the following comments:
\begin{itemize}
\item[$\checkmark$] If the tumor is too small then the treatment is stopped, otherwise, \\
\item[$\checkmark$] If $E\ge -\alpha r$, this is interpreted as {\em the tumor is not decreasing enough}, then the maximum drug intensity is used. \\
\item[$\checkmark$] If $E\le -r$, this is interpreted as {\em the tumor is decreasing fast enough} and the drug delivery is interrupted to privilege the patient health and to save drugs. \\
\item[$\checkmark$] The remaining hysteresis-like rule are used to define the control level over $(-r,-\alpha r)$ 
\end{itemize} 
To summarize the discussion regarding the definition of the feedback law, it comes out that the vector of control design parameter can be defined by:
\begin{eqnarray}
\theta_c:= \begin{pmatrix}
\beta & r & \alpha
\end{pmatrix}^T\in [1,\infty]\times [0,\infty]\times (0,1)\subset \mathbb{R}^{3}
\end{eqnarray} 
and more realistic set $\Theta_c$ to which these parameters can be defined by:
\begin{eqnarray}
\Theta_c:=\left\{1.05,2\right\}\times \left\{0.05,0.25,0.5,0.8\right\}\times \left\{0.1,0.5,0.8\right\} \nonumber \\
\ \label{defdeThetac} 
\end{eqnarray} 
which is a set of cardinality $24$. \ \\ \ \\ 
To this set of options, we have to add the extra parameters $N_T$, $\gamma$, $\gamma_c$, $D_1$ and $D_2$ that are involved in the definition of the therapy parameter $\theta$ [see (\ref{gftrtr})]. For this example, the current specific choices will be used regarding these design parameters:
\begin{itemize}
\item[$\checkmark$] The contraction factor $\gamma_c=0.1$ is fixed. 
\item[$\checkmark$] Duty cycle $\gamma\in \{0.3,0.5,0.8\}$
\item[$\checkmark$] Number of sub-periods $N_T=\{4,6\}$
\item[$\checkmark$] Available quantities of drugs: This is parametrized to be a quantized fraction of the maximum injectable quantity given the therapy duration $T$ and the duty cycle $\gamma$:
\begin{eqnarray}
D_i:=d\times \gamma T\bar u_i\qquad d\in \{0.25,0.5,0.75,1\} \label{defded} 
\end{eqnarray} 
where $\gamma T\bar u_i$ is the total amount of drug that is possible 
to inject given $T$, $\gamma$ and the maximal intensity bounds $\bar u_i$.  
\end{itemize}  
This new set of parameters is of cardinality $24$ which together with the set of control parameter leads to a total cardinality $$n_\Theta=24\times 24=576$$
In the sequel, the minimum number of circulating lymphocytes is taken equal to $C_{min}=5\times 10^7$. The bounds $\bar u_1=50$ and $\bar u_2=1$ involved in (\ref{defdeu1max}) and (\ref{defdeu2max})  are used. The total duration of the therapy is taken equal to $T=60$ days. The threshold involved in (\ref{defdethreshold}) below of which the treatment the drug injection is stopped is fixed in the sequel to $10^{4}$. The sampling period used to update the feedback is taken equal to $\tau=4$ hours. 
\ \\ \ \\  Before getting into the certification issue, Figure \ref{quatrescenaris} shows the results of the therapy using different set of therapy design parameters. These plots show how the choice of the design parameters systematically meet the constraint on the minimum level of circulating lymphocytes as it should be expected from the control law design while the contraction of the tumor and its intensity strongly depend on the parameter choices. All the scenarios start from the common initial state:
$$x_0= \begin{pmatrix}
5\times 10^9, 10^8, 0, 10^9, 0 , 0, T
\end{pmatrix} $$
\subsection{Generation of the model parameters sample}
\noindent Using the confidence level defined by $\delta=10^{-3}$, the precision level defined by $\delta=0.01$ and the cardinality $n_\Theta=576$ to compute the sample size $N$ using the expression (\ref{defdeNmini}), it comes out that the sample size if given by $$N=2155$$ 
Given the value of $n_\Theta=576$, this means that the computation of the best therapy design parameter $\theta$ needs $$\mbox{\rm Number of simulations = } 1,241,280$$
and since a single simulation of the feedback therapy over the therapy duration $T=60$ days costs approximatively $70\ \mu$ sec, it comes out that the whole computation of the parameters for a certified therapy can be done in approximatively $90$ seconds. \ \\ \ \\ 
Regarding the definition of the probability measure, as mentioned in section \ref{secappligeneral}, there are several ways to define how the true parameters spread around the nominal values given in Table \ref{tab2}. The one used hereafter to illustrate the methodology considers that each parameter $p_i$ of the model has a uniform probability over the following interval that includes the nominal value $p_i^{nom}$:
\begin{eqnarray}
p_i\in [\lambda_1,\lambda_2]\times p_i^{nom}\qquad (\lambda_1,\lambda_2)\in (0,1)\times (1,\infty)
\end{eqnarray}  
More precisely, if the pair $\lambda_1=0.6$ and $\lambda_2=1.8$ are used, this means that the probabilistic certification holds when each parameter can take with equal probability any value in the interval between $60\%$ the nominal value and $180\%$ of the nominal value.
\subsection{Validation}
\begin{figure*}
\begin{center}
\includegraphics[width=0.8\textwidth]{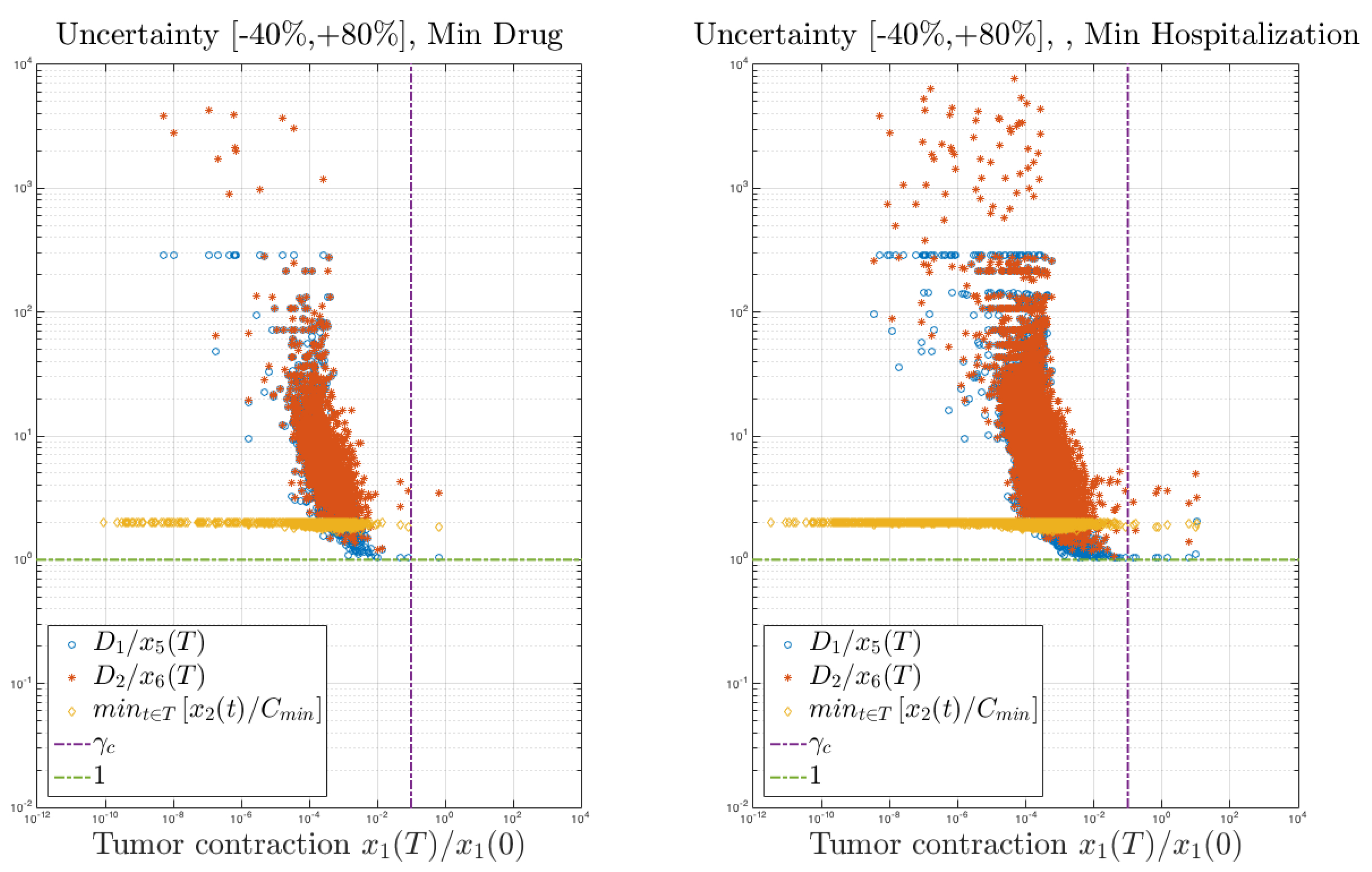}\ \\
(a) Uncertainty $[-40\%,+80\%]$ \vskip 0.5cm \ \\
\includegraphics[width=0.8\textwidth]{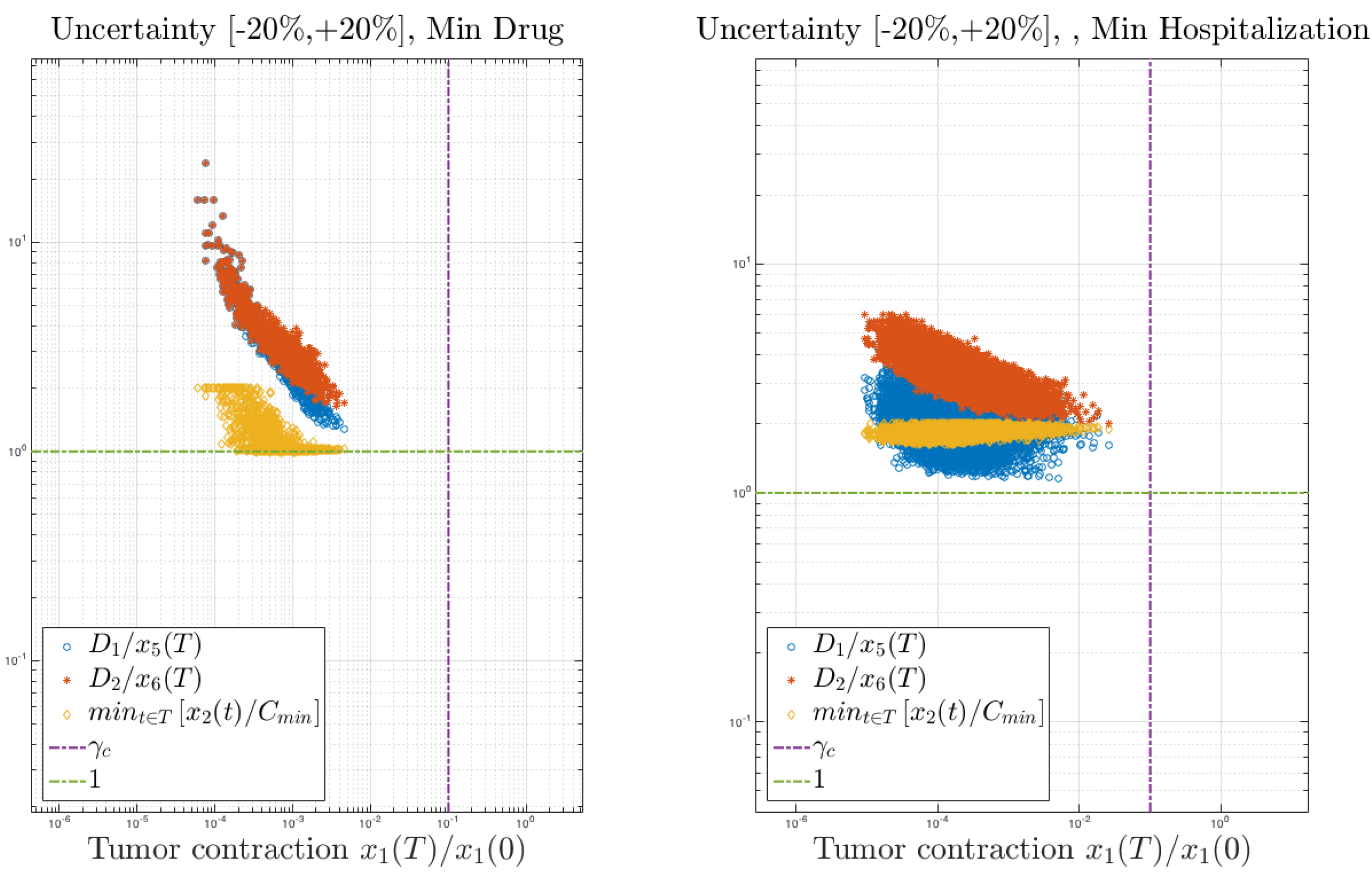}\ \\
(b) Uncertainty $[-20\%,+20\%]$ \vskip 0.5cm \ \\
\end{center} 
\caption{Validation of the certified optimal therapies over $5\times N\approx 11,000$ scenarios. The fact that almost all the dots appears in the upper left corner means that the contraction level ($\gamma_c=0.1$) is achieved, the health constraint is satisfied and that no more than the allowable is used.} \label{enfinvalidation} 
\end{figure*}

\noindent Several validation scenarios are proposed in this section depending on:
\begin{enumerate}
\item The level of uncertainties: Three couples of $(\lambda_1,\lambda_2)$ are used leading to three uncertainty levels: $[-10\%,+10\%]$, $[-20\%,+20\%]$ and $[-40\%,+80\%]$ which correspond to the pair $(\lambda_1,\lambda_2)$ given by: $(0.9,1.1)$, $(0.8,1.2)$ and $(0.6,1.8)$ respectively.\\
\item The criterion that is used to define the optimal parameter over the admissible set of values. namely, two criteria are used:\\ 
\begin{itemize}
\item[(a)] the minimization of the quantity of drugs. This is done by minimizing the parameter $d$ involved in the definition (\ref{defded}) of the quantity of drug available for the whole therapy.\\
\item[(b)] The minimization of the hospitalization periods. This is done by minimizing the parameter $\gamma$ which is the fraction of the treatment according to Figure \ref{fig_protocol}.  
\end{itemize} 
\end{enumerate} 
The objective is to show how the optimal therapy parameters are affected by these above paradigms leading to different but certified therapies over all possible realizations of the model's parameters. \ \\ \ 
\begin{table}
\begin{center}
\begin{tabular}{lcccc} \toprule
    {Uncertainties} & {Min drug} & {Min Hospitalization} \\ \midrule
    $[-10\%,+10\%]$  & $\begin{pmatrix}
\beta=1.05\cr r=0.25\cr \alpha=0.5\cr \gamma=0.8\cr N_T=4\cr \underline{\mathbf{d=0.25}}
\end{pmatrix}$ &       $\begin{pmatrix}
\beta=1.05\cr r=0.5\cr \alpha=0.8\cr  \underline{\mathbf{\gamma=0.3}}\cr N_T=4\cr d=0.75
\end{pmatrix}$ \\
\midrule 
    $[-20\%,+20\%]$  & $\begin{pmatrix}
\beta=1.05\cr r=0.25\cr \alpha=0.5\cr \gamma=0.8\cr N_T=4\cr \underline{\bf d=0.5}
\end{pmatrix}$ &       $\begin{pmatrix}
\beta=2\cr r=0.5\cr \alpha=0.5\cr \underline{\bf \gamma=0.3}\cr N_T=4\cr d=0.75
\end{pmatrix}$ \\
\midrule 
    $[-40\%,+80\%]$  & $\begin{pmatrix}
\beta=2\cr r=0.05\cr \alpha=0.5\cr \gamma=0.8\cr N_T=6\cr \underline{\bf d=0.5}
\end{pmatrix}$ &       $\begin{pmatrix}
\beta=2\cr r=0.25\cr \alpha=0.5\cr \underline{\bf \gamma=0.5}\cr N_T=6\cr d=1
\end{pmatrix}$ \\
 \bottomrule
\end{tabular}
\end{center} 
\ \\
\caption{Optimal therapy design for different level of model uncertainties and different cost function. } \label{tabrob} 
\end{table}
\ \\
Table \ref{tabrob} shows the optimal therapy design for these six different contexts. Several comments may help  for a better understanding of the results shown in this table:
\begin{enumerate}
\item Higher uncertainties implies higher values of $\beta$ as this parameter is used in (\ref{defdex3max}) to consider that the lower bound is $\beta C_{min}$ rather than the real lower bound $C_{min}$. In that sense, $\beta$ allows for a security margin on the constraint satisfaction.  \\
\item Minimizing drug and minimization hospitalization seem to have opposite effects, at least for the feedback design considered in the present paper. Indeed, higher values of $\gamma$ enable to reduce the oscillation in the tumor size that are induced by high periods of drug-free rest and hence use less total amount of drugs. \\
\item This last comment also explain why in the presence of high uncertainties, higher number of sub-periods becomes mandatory in order to reduce the drug-free rest periods. 
\end{enumerate} 
Note that some of these comments probably hold only for the feedback strategy adopted in the paper which is simply given here for the sake of illustration of the general certification methodology. \ \\ \ \\ 
Finally, Figure \ref{enfinvalidation} shows the validation of the certified strategies over a sample of scenario containing $5$ times more scenarios than those used in the sample of size $N$ for the optimization purposes. This corresponds to $5\times 2155=10775$ scenarios. The fact that almost all the dots belongs to the upper-left corner means that the tumor contraction  by at least a factor of $\gamma_c=0.1$ is achieved, the health constraint is satisfied and that the quantities of drug used during the therapy is lower than the allowable one. the fact that sometimes the quantity of drug used is $10$ times smaller than the available one comes from a specific combination of parameters in the interval that makes the decrease of the tumor possible without much drug injection. 

\section{Conclusions and Future Works} \label{secConclusion} 
\noindent In this paper, a general framework is proposed for the probabilistic certification of combined therapy of cancer under tumor contraction and health constraint. The proposed solution is based on the randomized method that enables to transform the standard robust worst-case approach by a tractable problem with probabilistic constraints. The general concepts introduced are illustrated in the specific case of combined immunotherapy/chemotherapy of cancer.\ \\ \ \\ The framework proposed in this paper can be used either to define the level of confidence that can be affected to a therapy with a given protocol and a given available quantity of drugs; or to determine what is the quantities of drugs and what is the protocol to be used in order to achieve a targeted level of confidence.  As such, the framework can be viewed as a decision making tool that enables different options to be compared based on reliable computation. \ \\ \ \\ 
As mentioned earlier, this contribution can be viewed as a starting point for a completely new approach to model-based control of tumors since it compensates for the oversimplifying character of population models by allowing high level of uncertainty on the value of the model's parameters. \ \\ \ \\ 
{\bf Acknowledgment} 
\noindent The author is grateful to professor T. Alamo (University of Sevilla) for the very fruitful discussions regarding the randomized approach. \\ \ \\  This work has been supported by the INSERM (Institut National de la Sant\'{e} et de la Recherche M\'{e}dical) projects CATS (Cancer Assisted Therapeutic Strategies). 
\bibliography{biblio_cancer_scenarios}
\bibliographystyle{plain}
\end{document}